\documentclass[twocolumn,showpacs,preprintnumbers,amsmath,amssymb]{revtex4}

\usepackage{graphicx}
\usepackage{dcolumn}
\usepackage{bm}
\usepackage[usenames]{color}

\newcommand{\change}[1]{\textcolor{blue}{#1}}

\begin{document}

\preprint{APS/123-QED}

\title{Prolonging assembly through dissociation : A self assembly paradigm in microtubules}

\author{Sumedha}
\author{Michael F Hagan}
\author{Bulbul Chakraborty}
\affiliation{Martin Fisher School of Physics, Brandeis University, Waltham,
MA 02454,USA}


\begin{abstract}

We study a one-dimensional model of microtubule assembly/disassembly in which GTP
bound to tubulins within the microtubule undergoes stochastic
hydrolysis.  In contrast to models that only consider a cap of
GTP-bound tubulin, stochastic hydrolysis allows GTP-bound tubulin
remnants to exist within the microtubule. We  find that these buried
GTP remnants enable an alternative mechanism of recovery from shrinkage, and enhances
fluctuations of filament lengths. Under conditions for which this alternative mechanism dominates, an increasing depolymerization rate leads to a decrease in dissociation rate and thus a net increase in assembly.

\end{abstract}

\maketitle

Microtubules are semiflexible polymers that serve as structural
components inside the eukaryotic cell and are involved in many
cellular processes such as mitosis, cytokinesis and vesicular
transport \cite{review1,review2,review3}. In order to perform these
functions, microtubules (MTs) continually rearrange through a process
known as dynamic instability (DI), in which they switch from a phase
of slow elongation to rapid shortening ( catastrophe), and from rapid
shortening to growth (rescue)\cite{review1}. The basic self-assembly mechanism underlying DI, assembly mediated by nucleotide phosphate activity\change{,} is omnipresent in biological systems. In this paper we
study a minimal non-equilibrium model of DI that shows enhanced assembly
with increasing depolymerisation rate. This provides a new paradigm of
self-assembly, which can occur only in non-equilibrium systems, and in
the context of DI, could explain some puzzling results about the influence of
proteins on MT assembly\cite{cassimeris}.

With recent advances in experimental techniques\cite{perez,gardner,howard-hyman},
it has become possible to quantify MT dynamics at nano-scale and,
thereby, provide more stringent tests of models. Models like ours can provide
insight into the non equilibrium phenomena of self-assembly and provide
a palette of scenarios. While it is established that GTP hydrolysis is essential
to DI, the mechanisms that underly DI are not fully understood. In this paper,
we study a minimal model of DI that involves stochastic { (or random)} hydrolysis
(SH), a mechanism that has received relatively little attention
compared to interfacial { (or vectorial)} hydrolysis (IH) that forms the basis of
cap models\cite{leibler,stukalin}. We study a particular SH
model\cite{chakraborty,antal}, which in contrast to a SH model that
takes into account all thirteen protofilaments
of a MT\cite{vanburen}, depicts the MT as a $1-d$ sequence
with rates that prescribe polymerization, depolymeriazation and
hydrolysis. The focus of our work is to relate the functioning of MT's to GTP
remnants that are characteristic of SH models.

MTs are formed by assembly of $\alpha-\beta$ tubulin dimers, which are
polar and impart polarity to MTs. MTs grow mainly from the end that
has exposed $\beta$ tubulin, and are composed of (typically) 13
linear protofilaments.\cite{review1}  While a free tubulin dimer has a GTP molecule bound to each monomer,
incorporation into a MT activates the $\beta$-tubulin monomer for
hydrolysis of its associated GTP. GDP-bound tubulin is less stable
within the MT lattice \cite{gebremichael} and hence a GDP-bound
tubulin at the tip of a MT has a higher rate of detachment
(depolymerization) than a GTP-bound tubulin. GTP hydrolysis is
essential to DI. Models in the IH class assume that all
hydrolysis occurs at a sharp interface between GDP-bound and GTP-bound
tubulins \cite{leibler,stukalin}, whereas in SH-based models,
hydrolysis occurs stochastically, anywhere in the MT \cite
{vanburen,chakraborty,flyvbjerg,antal,hill,margolin}.

In contrast to IH models, SH models lead to GTP-monomers being located throughout  the MT, with a
concentration that decays exponentially with  distance from the
growing end\cite{antal}. These models allow for a re-polymerization
mechanism that involves these GTP remnants, i.e as the MT depolymerises by detachment of GDP-tubulins,
the remnants get
exposed and the MT starts polymerising again.
Support for presence of GTP tubulins inside the MT has been provided
by recent experiments\cite{perez}.  The remnant-mediated re-polymerization leads to the 
possibility of extending activity through increased depolymerization rates.



\begin{figure}[!h]
\centering
\includegraphics[width=8.6cm]{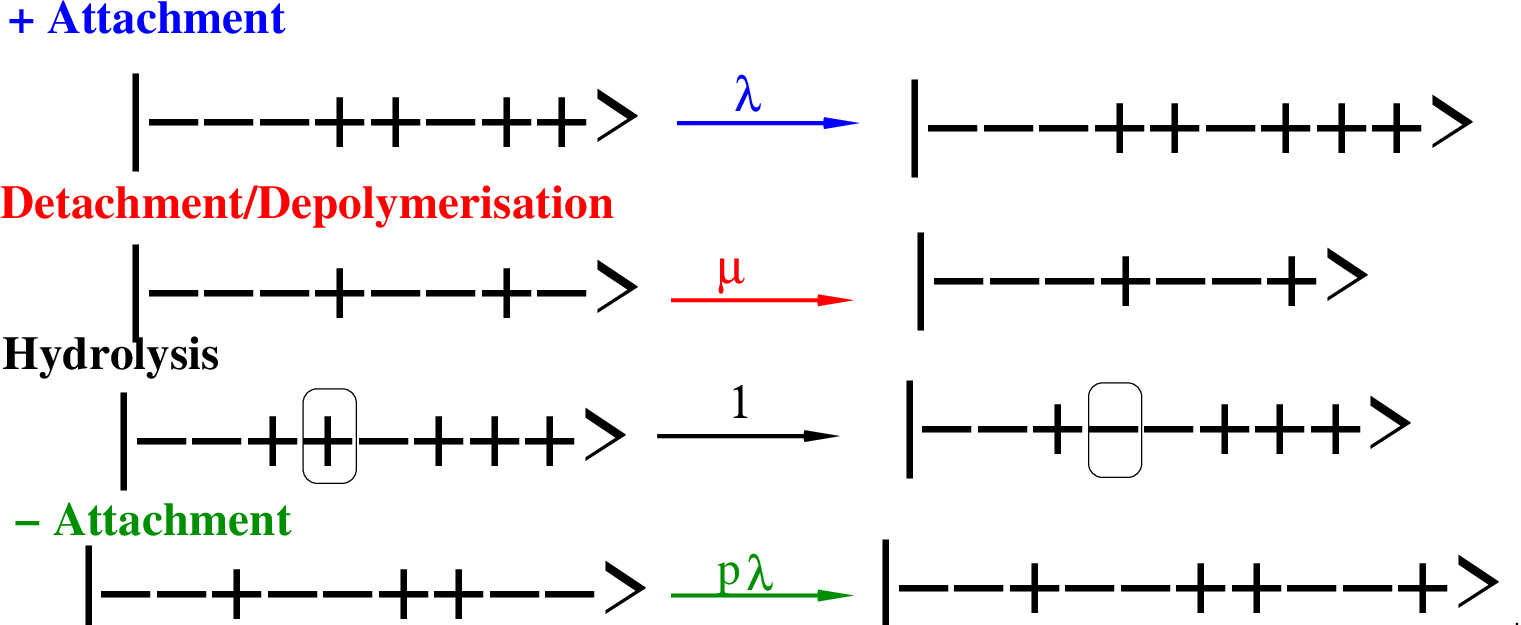}
\caption{\small{(Color online) Schematic of microtubule dynamics. We assume that all the activity 
occurs at the right end (denoted by $>$) of the MT.}}
\label{fig0}
\end{figure}

Our model\cite{chakraborty,antal} represents the MT by a linear
sequence of two species of monomers, which correspond to GTP-bound
tubulin (denoted by $+$ in rest of the paper) and GDP-bound tubulin
(denoted by $-$). We assume that the MT undergoes attachment and
detachment only at one end, which we call the growing end (sometimes
called the $+$ end in the literature). A MT evolves via the following rules (illustrated in
Fig. \ref{fig0}):
\begin{enumerate}
\addtolength{\itemsep}{-0.5\baselineskip}
\item \emph{$+$ attachment:} If the growing tip is a $+$ monomer, it grows
  with rate $\lambda$ by addition of a $+$ subunit.
\item \emph{Detachment:} A $-$ monomer at the growing end detaches
  with rate $\mu$, causing its shrinkage.
\item \emph{Hydrolysis:} With rate 1 any $+$ monomer in the MT can
  undergo hydrolysis to yield a $-$ monomer.
\item \emph{$-$ attachment:}  $+$ subunits could attach to a
  growing end with a $-$ monomer at the tip with rate $p \lambda$
  ($p\le 1$).
\end{enumerate}
A previous study of the model \cite{antal} for $p>0$, demonstrated  a
transition from a phase of bounded to unbounded growth  of the
MTs. The present study focuses on low $p$ and fluctuations in the
bounded growth region of the phase diagram.
The effect of remnants on dynamics of MTs is strongest for $p=0$ model,
which we call the {\it GTP remnants model} since the only mode of recovering from depolymerization 
is via remnants. Recent experiments indicate correlation between the presence
of remnants with events where the MT switches from shrinkage to growth. In particular, Perez et
al. \cite{perez} observe  GTP-bound tubulin within MTs and find that
the location of these remnants correlate to locations at which such 
events occurred during MT growth.

In the remnants model, if the number of GTP monomers  fluctuates to 0,
growth is no longer possible and the MT eventually shrinks
completely. Any process that exposes remnants promotes growth fluctuations and
makes the MT remain active for longer. In particular, increasing
$\mu$  at fixed $\lambda$ leads to longer times ($t_N$), and higher
maximum lengths ($L(t_N)$) of MTs before complete loss
of GTP. Beyond $t_N$ the MT undergoes a catastrophe, and
shrinks to zero in time of order of $L(t_N)/\mu$. The time $t_N$ can,
therefore be thought of as an activity time \footnote{We have considered full
dissociation of MT as catastrophe. This is different than the process of
detachment that occurs during the growth phase of MT}.

\begin{figure*}
\begin{tabular}{cc}
\includegraphics[width=8cm]{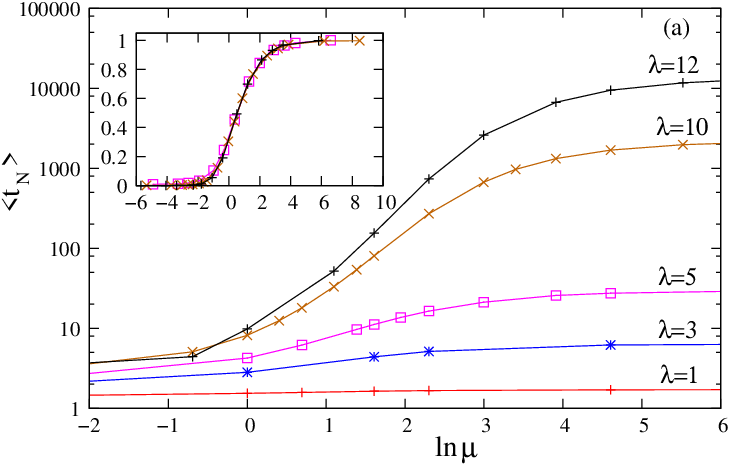} &
\includegraphics[width=8cm]{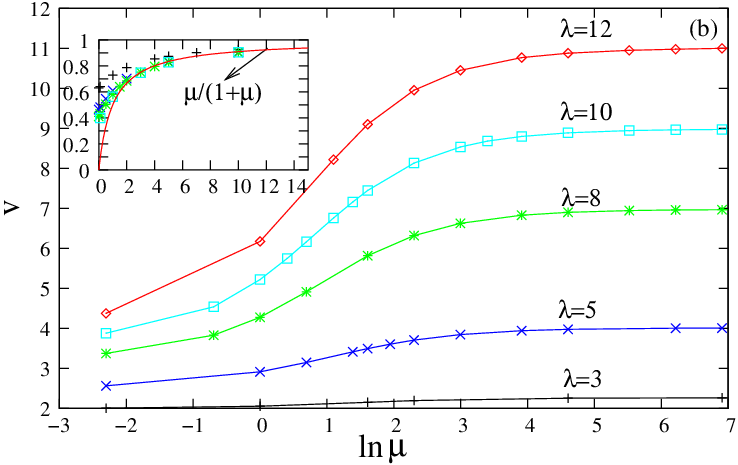}\\
\end{tabular}
\caption{\small{(color online){\it Left:} The average time at which the number of
    GTP-bound subunits goes to zero ($<t_N>$) as a function of $\ln
    \mu$. The inset shows a scaling collapse of the plots, with
    $y$-axis  scaled by $e^{\lambda}/\lambda$ and the $x$-axis
    displaced by $\lambda$, for  $\lambda=8,10,12$. {\it Right:} The
    average growth velocity  $(v=<L(t_N)>/<t_N>)$ for indicated values
    of $\lambda$ as a  function of  $\ln \mu$. In the inset, we have
    scaled $v$ for $\lambda=3,5,8,10,12$, and $\mu>1$, by $(\lambda-1)$  to illustrate the scaling
    collapse implied by Eq. \ref{eq6} for different values of $\lambda$. We find the scaled values lie on the $\mu/(1+\mu)$ curve. The scaling holds for $\lambda \geq 8$, with data for $\lambda=5$, showing deviations at small values of $\mu$.}}
\label{fig2}
\end{figure*}

Fig \ref{fig2} shows numerical results $<t_N>$ as a function of $\ln\mu$ for various values of
$\lambda$. It is clear that the activity time increases monotonically with increasing
$\lambda$ and $\mu$. For a given $\lambda$,  $<t_N>$ increases with increasing $\mu$, and eventually
saturates  at a value of the order of $\exp(\lambda)/\lambda$.
In Fig. \ref{fig2},  we also  show numerically obtained values  of the average growth
velocity$(v=<L(t_N)>/<t_N>)$ at time $t_N$. The velocity increases
with increasing $\lambda$ and $\mu$, and for a given $\lambda$,  it
saturates for large $\mu$.  As illustrated in  Fig. 2,
$v=(\lambda-1)\mu/(1+\mu)$  leads to good scaling collapse of the
data. Both of these features illustrate the increase in activity with increasing
{\it depolymerization} rate,  which is a hallmark of growth fluctuations initiated by the presence
of remnants.  As shown below, $-$ attachment events do not destroy this
signature for small $p$.

To allow experimental verifications of the predictions of the model at different
values of the system parameters, we obtain approximate analytical expressions for
distributions and averages of lengths etc for the model.
The probability distribution of total length $L(t)$ of the MT at time
$t$ follows  the following equation:
\begin{equation}
\begin{split}
\frac{d P(L,t)}{dt} =  \lambda (1-n_{0}(t)) [P(L-1,t)-P(L,t)] + \\ \mu
n_0(t) [P(L+1,t)-P(L,t)]
\end{split}
\label{eq3}
\end{equation}
where $n_0(t)$ is the probability of having a GDP at the tip of MT and
is given by:
\begin{equation}
\frac{dn_{0}(t)}{dt} = 1-n_{0}(t) - \mu n_{0}(t)P(+->,t)
\label{eq4}
\end{equation}
here $P(+->,t)$ is the conditional probability that, given the tip of the MT is a
$-$, the second last tubulin dimer is a $+$. A configuration with $|+->$ at the tip can 
be reached either by depolymerisation of $|+-->$ state or by hydrolysis of a $|++>$ state. 
The probability of having a $|+-->$ configuration further depends on $|+--->$ and so on. 
This is the reason why it is not possible to calculate $P(|+->,t)$ exactly for arbitrary 
values of $\lambda$ and $\mu$.

\begin{figure}[!h]
\centering
\includegraphics[width=8.6cm]{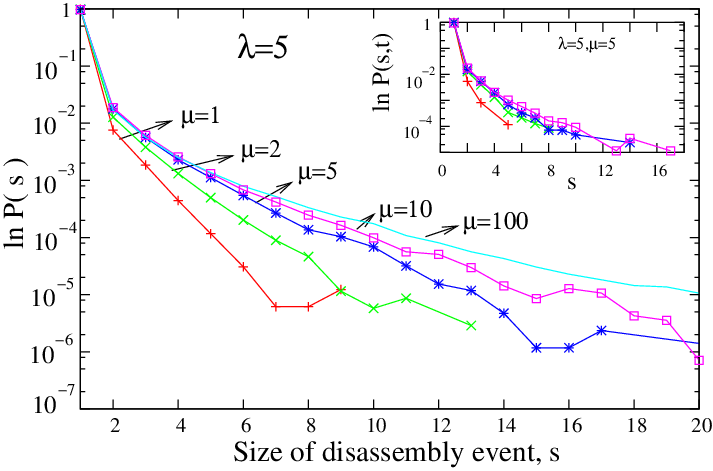}
\caption{\small{(color online) Distribution of size of disassembly events for $\lambda=5$ 
on a semilog plot}}
\label{figx}
\end{figure}

Let us first consider $\mu=0$  for arbitrary value of $\lambda$. In this limit,
there is only polymerization and hydrolysis that converts GTP to GDP, and $n_0(t) =
1-\exp(-t)$.  Substituting in Eq. \ref{eq4} leads to:
\begin{equation}
P(L,t) = e^{(-\lambda (1-e^{-t}))} \frac{(\lambda (1-e^{-t}))^L}{L!}
\label{eq2}
\end{equation}
The average length at any time $<L(t)>= \lambda
(1-\exp(-t))$. Similarly,  the average number of GTP-bound tubulins at
time $t$ is $<T(t)> =\lambda t \exp(-t)$.  Hence, the average time at
which the amount of GTP goes to zero (purely through hydrolysis) scales roughly as
$\ln(\lambda)$.

\begin{figure}[!htb]
\centering
\begin{tabular}{cc}
\includegraphics[width=4.4cm]{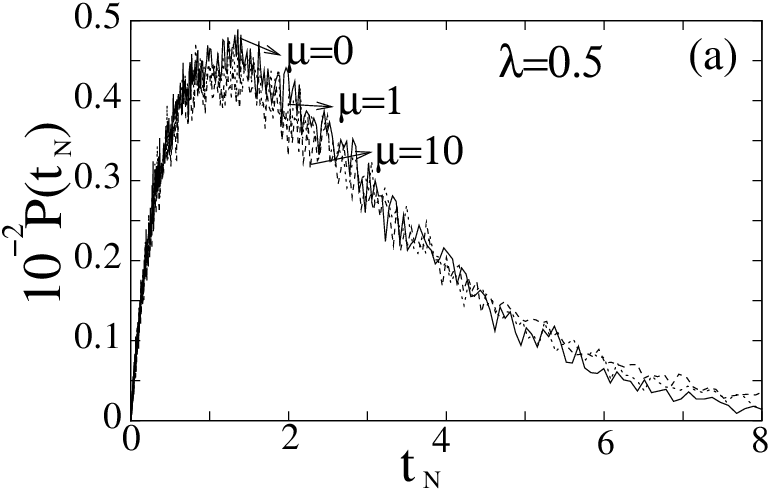} &
\includegraphics[width=4.4cm]{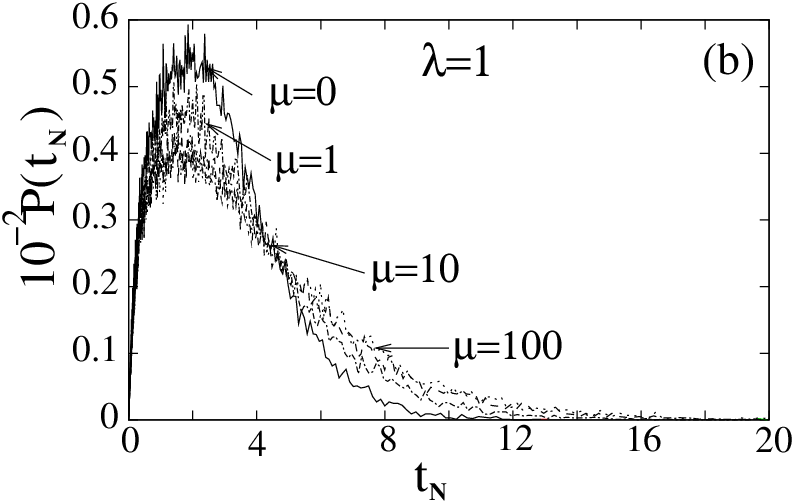}\\
\includegraphics[width=4.4cm]{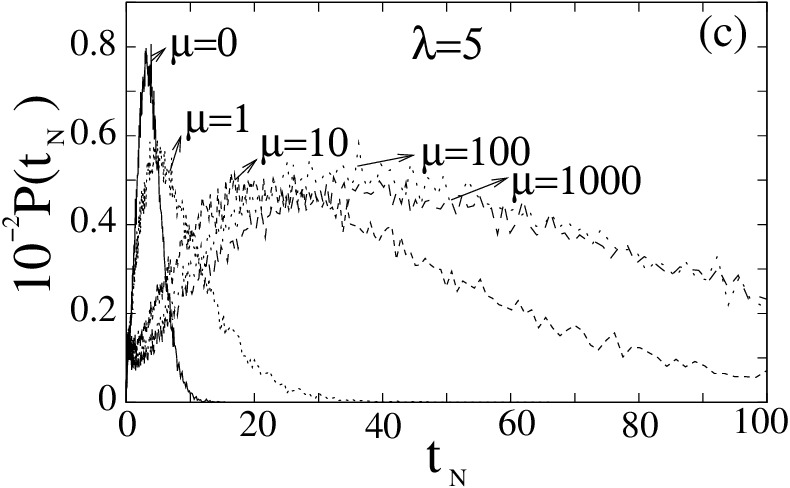}
& \includegraphics[width=4.4cm]{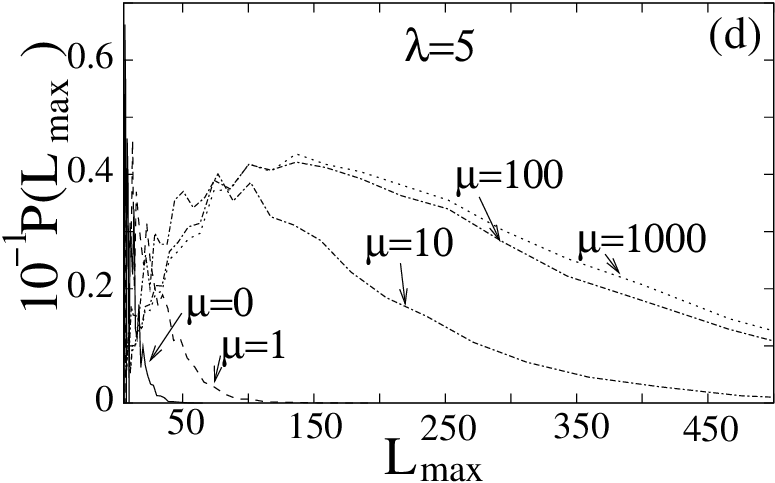}\\
\end{tabular}
\caption{\small{Color online {\bf (a)-(c)} The distributions of times at which the
    amount of GTP in the MT goes to zero$(t_N)$. The
    distribution shifts to the right with increasing $\mu$ until it
    saturates. (d) The distribution of maximum lengths for
    $\lambda=5$. The distributions of maximum lengths and $t_N$ have
    similar dependencies on $\mu$.}}
\label{fig3}
\end{figure}

Introducing a non-zero $\mu$ enables depolymerization, which dramatically changes assembly behavior by 
exposing remnants buried inside the MT to offer the possibility of a growth fluctuation. It is difficult 
to obtain analytic solutions to Eq. \ref{eq4} at finite values of $\mu$ because of the coupling between 
$n_0$ and $P(+->,t)$, but numerical results demonstrate an  exponential dependence of
$t_N$ on $\lambda$ for large $\mu$ (indicated by the scaling collapse
in Fig.  \ref{fig2}). As seen in Fig. \ref{fig2}, $<t_N>$ increases
monotonically  from a value of order $O(\ln \lambda)$ to a value of
order  $\exp(\lambda)$ as we increase $\mu$ from $0$ to $\infty$.
The distributions of $t_N$ and $L(t_N)$ for various $\lambda$ and
$\mu$ (Fig. \ref{fig3}) broaden with increasing $\mu$, reaching
asymptotic forms for $\mu >> \lambda$.   The increase of fluctuations,
indicated by these broadening distributions, is a consequence of
remnants.

The coupling between $n_0$ and $P(+->,t)$ is the hurdle in obtaining analytical results with 
finite depolymerization rates. {\it But we can still try to make an indirect estimate as $P(+->,t)$ 
should be proportional to probability of disassembly events of size 1}. For finite values of $\lambda$ 
and $\mu$, we found numerically that most disassembly events in the growth phase 
involved $O(1)$ sites.  Fig. \ref{figx} shows the distribution of the size of disassembly 
events in the growth phase for $\lambda=5$ and $\mu=1,2,5,10$, and $100$, obtained by by 
averaging over 10000 growth events. {\it (by growth phase we mean $t<t_N$). To obtain the 
distribution in the growth phase we looked at all the disassembly events except the final 
complete disassembly}. The distribution is exponential for all values of $\mu$ and the average 
size of a disassembly event varied from $1.013$ to $1.06$ as $\mu$ was changed from $1$ to $100$. 
{\it This suggests that growth happens predominantly because the $-$ at the tip detaches before the 
hydrolysis of $+$ next to the tip. Infact we expect that the competition between these two events determines 
the value of $t_N$. There are ofcourse rare event where larger size disassembly 
events occur during the growth phase, but we expect them not to influcence the dynamcis of the 
MT significantly. Hence, we take $P(|+->,t) \approx 1$ as a good approximation for our model MT in 
its growth phase.}  This approximation is exact in the limit of  $\lambda \rightarrow \infty$, 
$\mu \rightarrow \infty$ but based on the numerics we adopt it for finite values of the rates. 
Setting $P(|+->,t)=1$, in Eq. \ref{eq4} we obtain,


\begin{equation}
n_0(t)=\frac{1}{1+\mu} (1-\exp^{-t(1+\mu)})
\end{equation}
Substituting in Eq. \ref{eq3} leads to ,
\begin{equation}
P(L,t)= (\lambda)^{L/2} I_L \left(\frac{2 \mu t
  \lambda^{1/2}}{1+\mu}\right) \exp\left( -\frac{\mu
  (\lambda+1)t}{1+\mu}\right)
\label{eq5}
\end{equation}
where $I_L(x)$ is the modified Bessel function of first kind and
Eq. \ref{eq5} leads to:
\begin{equation}
 <L(t)> = \frac{\mu(\lambda-1)}{(1+\mu)}t
\label{eq6}
\end{equation}
Eq. \ref{eq6} implies that the $\mu$-dependence of the growth velocity curves for different values of $\lambda$ can be scaled on to each other. 
In Fig. \ref{fig2}(right) we have plotted the velocity obtained for a range of  $\lambda$ and $\mu$ values. 
As shown in inset, on scaling the average velocity in the growth regime with $(\lambda-1)$ we get a scaling 
collapse for different $\lambda$'s and the scaled curve lies on the $\mu/(1+\mu)$.

The equation obtained with the approximation, $P(+->,t)=1$ describes only the growing phase. It cannot 
describe the dissociation of the MT that occurs beyond the time $t_N$.   Strictly speaking, the results 
from Eq. \ref{eq6} apply only for times much shorter than $t_N$.  The numerics, however,  indicate that 
the approximation remains valid even pretty close to $t_N$.  Although  Eq. \ref{eq5} is based on the 
assumption $P(+->,t)=1$, which is exact only in the limit of diverging values of $\mu$ and $\lambda$, 
our numerical results indicate that fluctuations in the growth phase are well described by Eq. \ref{eq5}. 
In fact, for a given $t$, the  distribution of
lengths (Eq. \ref{eq5}) broadens with $\mu$ in a manner similar to that
seen in the simulations (Fig.(4)). The variance of distribution of
$P(L,t)$ from Eq. \ref{eq5} comes  out to be
$\frac{\mu (\lambda+1)}{1+\mu} t$.

Similarly, one can solve for the distribution and mean of $T(t)$. The average value 
$<T(t)>$ follows the following equation:
\begin{equation}
\frac{d <T(t)>}{dt} = \lambda (1-n_0(t))-<T(t)>
\end{equation}
Assuming $P(+->,t)=1$, we get:
\begin{equation*}
 <T(t)> = \frac{\lambda \mu}{1+\mu} - \frac{\lambda exp(-(1+\mu)
   t)}{\mu (1+\mu)}-\frac{\lambda (\mu-1) exp(-t)}{\mu}
   \label{blah}.
\end{equation*}
This equation also matches the simulation results, where we found that
the average amount of GTP in the MT during the growing phase
fluctuates around $\frac{\lambda \mu}{(1+\mu)}$.

We do not have a good understanding of why the $P(+->,t)=1$ describes the dynamics of our model over a 
large range of parameters. If we, however, assume the approximation to be valid, we can make a number 
of other predictions, which should apply to the growing phase, $t <<\langle t_N \rangle$. For example, 
the average number of disassembly events at time $t$ in which the MT switches from a growing to a 
shrinking phase, is given, to leading order  by $<C(t)> = \frac{\mu^2 t}{(1+\mu)^2}$ and the average 
number of GTP islands in the MT is predicted to be: $\frac{\lambda \mu(1-\exp(-2t))}{2 (1+\mu)}$.

\begin{figure}[!h]
\centering
\includegraphics[width=8.6cm]{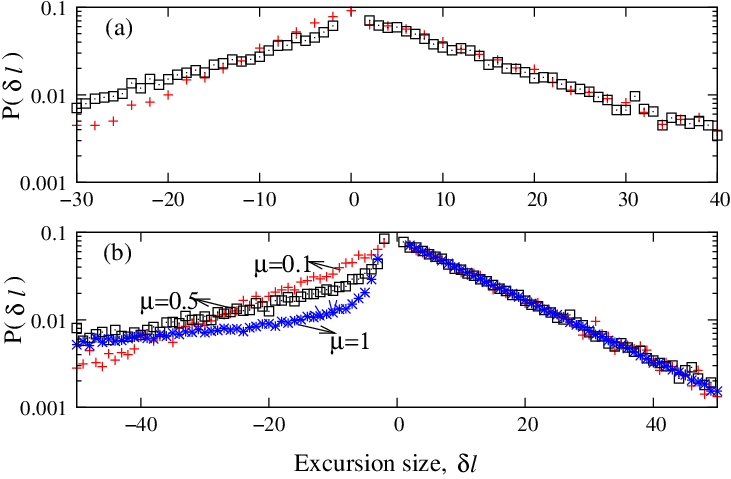}
\caption{\small{(color online){\it Top:} Comparison of
    excursions($\delta \it l$) predicted
    by the GTP remnants model ($\boxdot$) with $\lambda=12;p=0;\mu=0.1$, and
    experimental measurements ($+$) \cite{gardner}($\delta \it l$ is measured 
    in nm in the experiment). {\it Bottom:} The
    distribution of excursions with $\lambda=12$;$p=0$ for $\mu=0.1,0.5$ and
    $1$. While growth excursions have the distribution
    $\exp(-\delta {\it l}/\lambda)/\lambda$, the distribution of shortening
    excursions changes from being exponential to non exponential and broader
    distributions with increasing $\mu$.}}
\label{fig1}
\end{figure}

\begin{figure*}
\begin{tabular}{cc}
\includegraphics[width=8 cm]{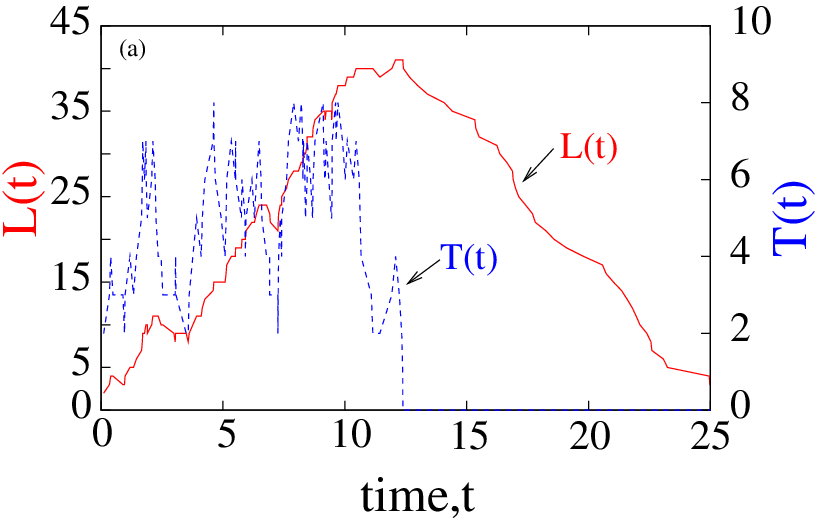} &
\includegraphics[width=8 cm]{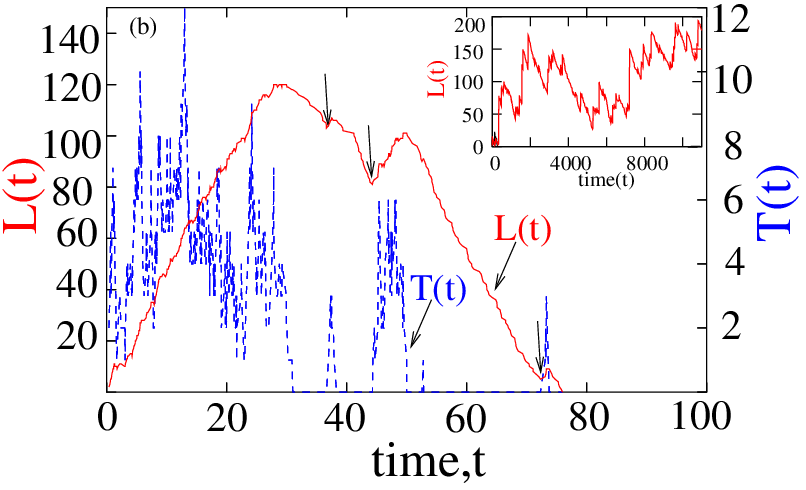}\\
\end{tabular}
\caption{\small{(Color online)MT length $L(t)$ (solid lines) as a function of time for
    $\lambda=5,\mu=4$ and $p=0$ ({\it Left}); $p=0.01$({\it
      Right}). The dotted lines show the total amount of GTP(T(t)) as a
    function of time for the same runs. Arrows in the second plot
   indicate the $-$ attachment events due to $p>0$. The inset shows MT length $L(t)$
    for $\lambda=12$,$\mu=0.1$ and $p=0.0007$. This trajectory looks qualitatively similar 
to the ones observed experimentally, and these growth phase fluctuations are different from 
the rapid shrinkage and slow growth observed in catastrophes and rescues \cite{gardner}}}
\label{fig4}
\end{figure*}

All of the predictions presented above, reflect the sensitivity of dynamical
properties to the depolymerization  rate $\mu$, and are a fingerprint of
remnants. Experimental tests of  these predictions can, therefore,
provide insight into the nature of  hydrolysis and polymerization-deploymerization mechanisms
in DI.



Recent experiments monitored\cite{gardner} the distribution of lengths
of growing and shortening excursions within the growth phase
in {\it in vitro} systems of MTs. These experiments were able to
resolve fluctuations at the monomer level, and the distributions
were found to be exponential. In simulations of our model we find
that for small values of $p$ the growth excursions are independent
of $\mu$ and can be fitted well by $\exp(-i/\lambda)/\lambda$
($i$ is the length of the excursion) and the distribution
of shortening excursions broadens with $\mu$ (Fig. \ref{fig1}).

We can fit the experimental data on the distribution of excursions
\cite{gardner} with our model by taking $\lambda=12;\mu \approx
0.1;0 < p \le 0.001$ (Fig. \ref{fig1}).
We have also plotted a typical trajectory with these values of the
parameters in the inset of Fig. 6. For these low values of $p$
obtained from the fits, GTP-remnants are the dominant source of 
recovery from negative growth events. These rates though 
representative for this 1-D model, correspond to effective rates for order 
13 subunits and thus are not quantitatively comparable to the 
rates for a thirteen protofilament MT \footnote{Note that growth phase shortening 
excursions are distinct from rapid shortening, which is more than an order of 
magnitude faster and typically persists for micrometers than nanometers.}.

The above analysis was restricted to  $p=0$  in order to highlight the
effect of remnants. As  $p$ is increased,
the dynamics changes from cessation of negative growth primarily due to remnants at small
$p$ to non-remnant attachment events that are also present in IH models at $p
\simeq 1$.  Fig. \ref{fig4} shows the time trace for MT length for a
representative run for $p=0$ and $p=0.01$.  Analysis of these
trajectories shows that a small, non-zero value of $p$ introduces rare 
$-$ attachment events (indicated by arrows in the figure).  These events
change the overall length of MT, but the statistics of positive and negative 
growth excursions remain similar to $p=0$. Measurement of these
statistics is possible in experiments such as the one analyzed above, 
and should provide tests of the remnant-induced  mechanism of growth fluctuations.

Our model can be easily extended to accomodate more detailed features
of MTs while keeping the basic mechanism of remnant-induced growth.
Parameters can be obtained from simulations by systematically mapping
simulations of microscopic models to our effective model.
For example, spatially varying hydrolysis rates due to the structure
of MTs\cite{vanburen} can be modeled by quenching some GTP-bound sites
in our 1-d model. Similarly, the effect of motors that mechanically
depolymerize MTs without dependence on GTP-states \cite{howard} can be
modeled by assuming a depolymerizing rate $\mu$ for both GTP and
GDP-bound tubulins.

Preliminary studies of the model with quenched
disorder indicate that, although the time for which MT grows changes
and there is a transition to unbounded growth as a function of percent
of quenched sites, the distribution of excursions and velocity of
growth remains unchanged, and therefore is a robust feature of the
remnant model.  Studies of the model mimicking motors also indicate
that the basic features of the remnant model remain unchanged as long
as $\lambda > \mu+1$.

To summarize, we have studied the role of GTP-remnants in MT
dynamics, and shown that remnants  give rise to features of DI that
are very different from IH models that have no remnants. Some particularly notable features
are:  1) the average catastrophe time increases with depolymerization rate.  2) the
distribution of MT lengths and time of growth depends on $\lambda$ and $\mu$,
broadening as $\lambda$ and $\mu$ are increased. 3) The velocity of growth,
besides depending on the free tubulin concentration( through
$\lambda$), also depends on depolymerisation and hydrolysis rates.
Similar behaviour was reported by Cassimeris et al \cite{cassimeris}. They found
that increasing concentration of XMAP resulted in increase of both depolymerisation
rate and growth velocity.
These features are robust, and with recent progress in experimental
techniques \cite{perez,gardner}, should provide tools for resolving the
mechanism of hydrolysis inside a MT. In conclusion, a minimal model of 
MT dynamics where cessation of negative growth is dominated by GTP remnants 
leads to strong spatial structure-dynamics connection.  The simplicity of our model 
allows us to make analytic predictions that can be tested experimentally, and provide sensitive
tests for the remnant-mediated mechanisms of growth in MT dynamics, both in-vivo and in-vitro.  Interestingly, at
the same tubulin concentration, MTs exhibit much higher growth rates \emph{in-vivo} in
comparision to \emph{in-vitro}\cite{review2,cassimeris1}.  In a broader context, the model
illustrates a new paradigm of non-equilibrium self assembly where assembly is promoted through
depolymerization.

We thank the authors of \cite{gardner} for providing us with their
data for MT excursions in-vitro. S and MFH acknowledge support by  NIH
grant  R01AI080791, and S, MFH and BC were supported in  part by the
Brandeis NSF-MRSEC.


\begin{thebibliography}{99}
\bibitem{review1} E. Karsenti,F. Nedelec and T. Surrey, Nature Cell
  Biology {\bf 8}, 1204 (2006).
\bibitem{review2} A. Desai and T. J. Mitchison, Annu. Rev. Cell. dev.
  Biol. {\bf 13},83-117(1997).
\bibitem{review3} J. Howard and A. A. Hyman, Nature {\bf 422}
  753(2003).
\bibitem{cassimeris} L Cassimeris, N. K Pryer and E. D. Salman, The journal of
cell biology {\bf 127} 985-993 (1994).
\bibitem{perez} A.Dimitrov, M. Quesnolt, S. Moutel, I. Cantaloube,
  C. Pous and F. Perez, Science {\bf 322} 1353-1356(2008).
\bibitem{gardner} H. T. Schek, M. K. Gardner, J. Cheng, D. J. Odde and
  A.J. Hunt, Current Biology {\bf 17} 1445-1455 (2007).
\bibitem{howard-hyman}, J. Howard and A. A. Hyman, Nature
Reviews Molecular Cell Biology, {\bf 10} 569-574 (2009).
\bibitem{leibler} P. M Bayley, M J Schilstra and S. R. Martin, J. of Cell Science {\bf 95}, 33(1990);
 M. Dogterom and S. Leibler, Phys. Rev. Lett.  {\bf
  70}, 1347 (1993); H. Flyvbjerg, T.E. Holy and S. leibler,
  Phys. Rev. Lett. {\bf 73}, 2372 (1994)
\bibitem{stukalin} D. Vavylonis, Q. Yang and B. O. Shaughnessy, PNAS {\bf 102} 8543(2005);
E.B. Stukalin and A. B. Kolomeisky, Bio. Phys. J. {\bf 90},  2673 (2006).
\bibitem{chakraborty} B. Chakraborty and R. Rajesh, unpublished.
\bibitem{antal} T. Antal, P.L. Krapivsky, S. Redner, M. Mailman and
  B. Chakraborty, Phys. Rev. E {\bf 76}, 041907 (2007);T. Antal,P.L. Krapivsky
and S. Redner, J. Stat. Mech.,L05004 (2007).
\bibitem{vanburen} V. Van Buren, D. J. Odde, and L. Cassimeris, PNAS
  {\bf 99} 6035-6040 (2002);V. van Buren, L. Cassimeris and D. J. Odde,
Biophysical Journal, {\bf 89} 2911(2005).
\bibitem{gebremichael} Y. Gebremichael, J. W. Chu and G. A. Voth,
  Biophysical Journal {\bf 95} 2487-2499 (2008).
\bibitem{hill}T.L. Hill, Biophys. J. {\bf 49}, 981(1986)
\bibitem{flyvbjerg} H. Flyvbjerg et al, Phys. Rev. E {\bf 54}, 5538
  (1996).
\bibitem{margolin} G Margolin, I. V. Gregoretti, H. V. Goodson and M. S. Alber,
Phys. Rev. E {\bf 74}, 041920 (2006).

\bibitem{howard} V. Varga, J. Helenius, K. Tanaka, A. A Hyman,
  T. U. Tanaka and J. Howard, Nature Cell Biology {\bf 8}, 957-962
  (2006).
  Biophysical Journal {\bf 83},1809-1819 (2002).
\bibitem{hunter} A. W. Hunter and L. Wordeman, Journal of Cell Science
  {\bf 113},4379-4389 (2000).
\bibitem{cassimeris1} L Cassimeris, Cell Motil. Cytoskelet. {\bf 26}
  275-281 (1993).

\end{thebibliography}
\end{document}